 \documentclass[oldversion]{aa}
\usepackage{graphicx}
\usepackage{longtable}
\usepackage{txfonts,color, natbib}

\begin{document}
\newcommand{\mps}{m\,s$^{-1}$}
   
   \title{Large-scale horizontal flows in the solar photosphere}

   \subtitle{V. Possible evidence for the disconnection of bi-polar sunspot groups from their magnetic roots}

   \author{M. \v{S}vanda
          \inst{1,2,3}
          \and
          M. Klva\v{n}a
          \inst{2}
          \and
          M. Sobotka\inst{2}
          }

   \offprints{M. \v{S}vanda}

   \institute{Max-Planck-Institut f\"ur Sonnensystemforschung, Max-Planck-Strasse 2, D-37191, 
              Katlenburg-Lindau, Germany\\
              \email{svanda@mps.mpg.de}
         \and     
              Astronomical Institute, Academy of Sciences of the Czech Republic (v.~v.~i.), Fri\v{c}ova 298,
              CZ-25165, Ond\v{r}ejov, Czech Republic\\
              \email{mklvana@asu.cas.cz, msobotka@asu.cas.cz}
	 \and
	      Astronomical Institute, Faculty of Mathematics and Physics, Charles University, V~Hole\v{s}ovi\v{c}k\'ach~2, Prague, CZ-18200, Czech Republic\\
             }
  \date{Received 5 May 2009 / Accepted 10 Aug 2009}
  \abstract{In a recent paper (\v{S}vanda et al., 2008, A\&A 477, 285) we pointed out that, based on the tracking of Doppler features in the full-disc MDI Dopplergrams, the active regions display two dynamically different regimes. We speculated that this could be a manifestation of the sudden change in the active regions dynamics, caused by the dynamic disconnection of sunspots from their magnetic roots as proposed by Sch\"{u}ssler \& Rempel (2005, A\&A 441, 337). Here we investigate the dynamic behaviour of the active regions recorded in the high-cadence MDI data over the last solar cycle in order to confirm the predictions in the Sch\"{u}ssler's \& Rempel's paper. We find that, after drastic reduction of the sample, which is done to avoid disturbing effects, a large fraction of active regions displays a sudden decrease in the rotation speed, which is compatible with the mechanism of the dynamic disconnection of sunspots from their parental magnetic structures. }

  \titlerunning{Large-scale horizontal flows in the solar photosphere V: A disconnection of bi-polar sunspot groups}

   \maketitle

\section{Introduction}

The dynamics of active regions in the solar photosphere and in the close sub-photospheric layers is closely related to the solar dynamo process. The magnetic flux emerges from beneath the surface and forms sunspots and other active phenomena. At some point, the magnetic field starts to disperse, the sunspots disappear, and the active region decays, forming the surge-like structures of the trailing polarity expanding to the solar poles \citep{1965ApJ...141.1502B}, where they contribute to the solar field reversals \citep[see e.g.][]{1989Sci...245..712W}. 

The measurements of the dynamic behaviour of active regions can bring forth new insights into what is happening to the magnetic field under the surface. Many authors have studied the variations in the rotation speed of sunspots. The rotation of the sunspots in relation to their morphological type was studied by e.g., \citet{1986AA...155...87B}, who found that more evolved types of sunspots (E, F, G, and H types) rotate slower than less evolved types. \citet{2004SoPh..221..225R} investigated the Greenwich Photoheliographic Results for the years 1874--1976 and found clear evidence for the deceleration of the sunspots in the photosphere with their evolution. \citet{2002aprm.conf..427H} found that the leading part of a complex sunspot group rotates about 3~\% faster than the following part. The dependence of  the rotation rate of sunspots on their size and position in the bi-polar region was investigated by \citet{1994SoPh..151..213D}. Their results showed that smaller spots rotate faster than large ones. The authors explained the observed behaviour through a subtle interplay between the forces of magnetic buoyancy and drag, coupled with the Coriolis force acting on rising flux tubes.

The interpretations used in the explanation of the observed dynamics of active regions often involve the deep anchoring of the magnetic structures. From helioseismic inversions, we know that throughout the convective envelope, the angular velocity contours at mid-latitudes are nearly radial. Near the surface, at the top of the convection zone, there is a layer of a large radial shear in the angular velocity. At low and mid-latitudes, there is an increase in the rotation rate immediately below the photosphere that persists down to $r \sim 0.95~R_\odot$ (where $r$ is the radial coordinate). The angular velocity variation across this layer is roughly 3~\% of the mean rotation rate, and according to the helioseismic analysis of \citet{2002SoPh..205..211C}, the angular velocity $\omega$ decreases within this layer approximately as $r^{-1}$. At higher latitudes, the situation is less clear. The changing depth of the magnetic roots, e.g., by the rising motion, could explain the deceleration of the sunspots during their evolution. 

One of the important properties observed in the bi-polar sunspot groups is their tilt with respect to the zonal direction \citep{1919ApJ....49..153H}. This tilt is believed to be generated by the Coriolis force, which acts to twist the flux-loop as it rises through the convection zone. The obvious question is, what happens to the tilt when the flux eruption has ceased, because this implies that the twisting Coriolis force no longer exists? If the magnetic field remains connected to the deep toroidal field at the base of the convection zone, then the tilt should slowly diminish through the action of the magnetic tension. However, this behaviour is not observed, which challenges the solidity of the connection to the deep magnetic field. 

A number of theories have been suggested \citep{1994ApJ...436..907F,2005AA...441..337S} indicating that bi-polar magnetic regions may become disconnected from their magnetic roots and form an isolated island-like feature. The mechanism is based upon the buoyant upflow of plasma along the field lines. Such flows arise in the upper part of a rising flux loop during the final phase of its buoyant ascent towards the surface. The combination of the pressure build-up by the upflow and the cooling of the upper layers of an emerged flux tube by radiative losses at the surface leads to a progressive weakening of the magnetic field at depths of several Mm. When the field strength has become sufficiently low, convective motions ablate the flux tube into thin, passively advected flux fragments. The mechanism provides for a dynamical disconnection of the emerged part from its parental magnetic structure. This instant should be observed as a change in the dynamic regime, as the ``floating island'' does not reflect the deep dynamics any longer. We note that even in the models which assume that the surface magnetic activity is a consequence of the shallow dynamo action \citep[e.g.][]{2009SoPh..255....3S}, the separation of the surface local magnetic field from the close subsurface one is required, in order to allow the magnetic flux to disperse towards the solar poles. The model of \cite{2009SoPh..255....3S} however does not provide the predictions for the systematic dynamics of active regions in the photosphere.

In the recent paper of \cite{2008AA...477..285S} we found that sunspots in the equatorial region seem to show two different dynamical regimes. In one regime, the \emph{fast regime}, the sunspots displayed an almost constant velocity of 1910$\pm$9~\mps{}. Furthermore, the sunspots embodying this regime were of a younger type. The second group, the \emph{scattered group}, contained mostly old and perhaps recurrent sunspot groups. In the velocity, this group showed a large scatter with the mean speed of 1850~\mps. At that time we did not have a tool to follow a particular region on the Sun for any time of interest, therefore we concluded that the existence of the above-mentioned regimes is compatible with the theory of dynamic disconnection. Furthermore, the fact that all the young sunspots we observed depicted almost the same speed, led us to speculate that this behaviour is related to the assumption that the magnetic field emerges from a similar depth in the convection zone, in line with deep anchoring. Following the approximation of the radial rotation profile in \citet{2002SoPh..205..211C}, the radius where the rotation corresponds to 1910$\pm$9~\mps{} is roughly 0.946$\pm$0.008~$R_\odot$.
 
In the past year, we have developed a tool for the selection of the active regions in the Michelson Doppler Imager \citep[MDI;][]{1995SoPh..162..129S} magnetograms. From the previous studies we possess the datasets that cover all the MDI \emph{Dynamics campaigns} in years 1996--2006, i.e., 502~days of high quality observations. In each of these days, two full-disc 24-hours-averaged horizontal velocity maps, with an effective resolution of 60\arcsec{} sampled by 12 hours, were calculated. This allowed us to study the evolution of the active region dynamics recorded in our velocity datasets, in particular the phenomenon of the dynamic disconnection from the magnetic roots.

\begin{figure*}[!t]
\centering
\resizebox{0.45\textwidth}{!}{\rotatebox{270}{\includegraphics{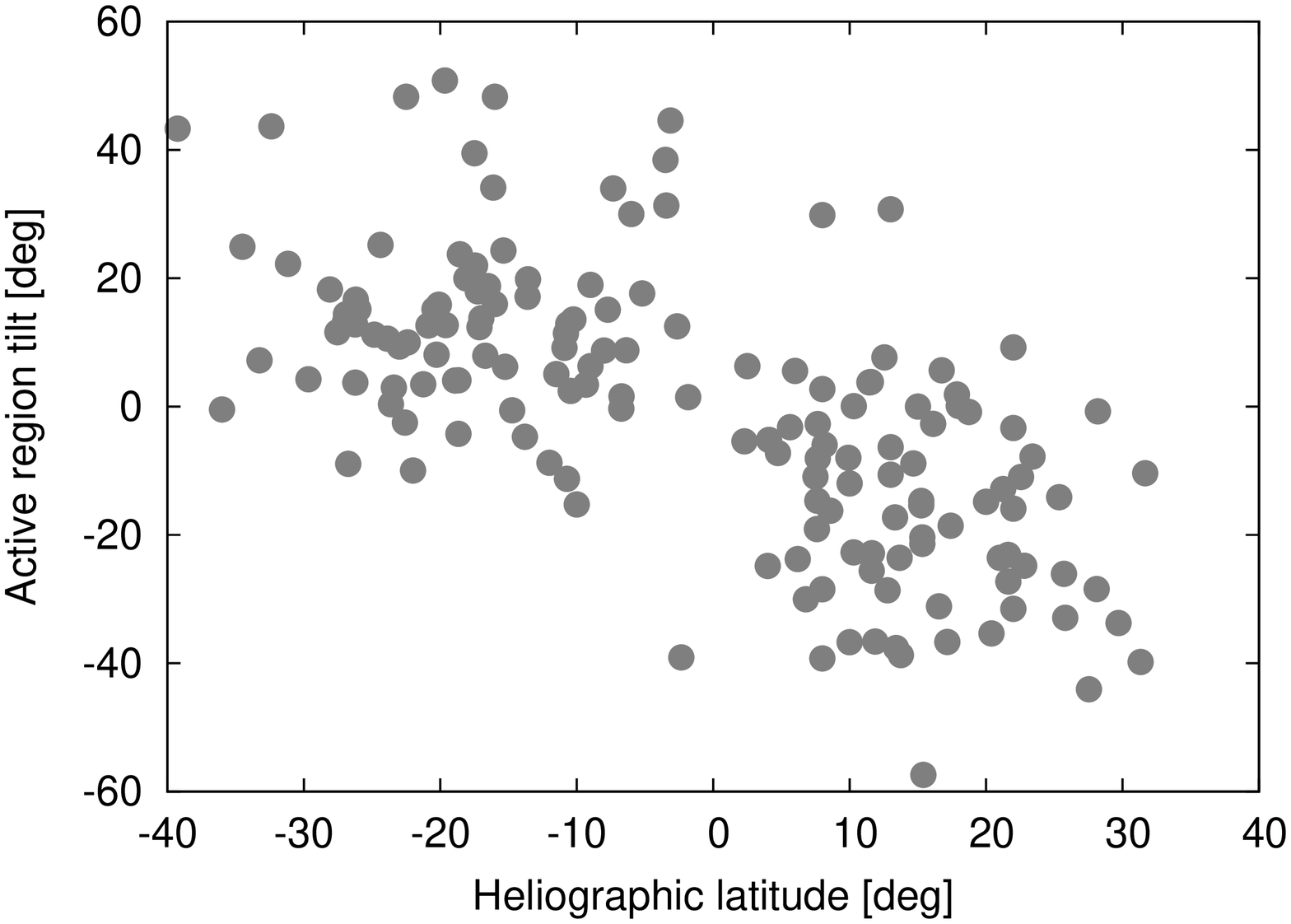}}}
\resizebox{0.45\textwidth}{!}{\rotatebox{270}{\includegraphics{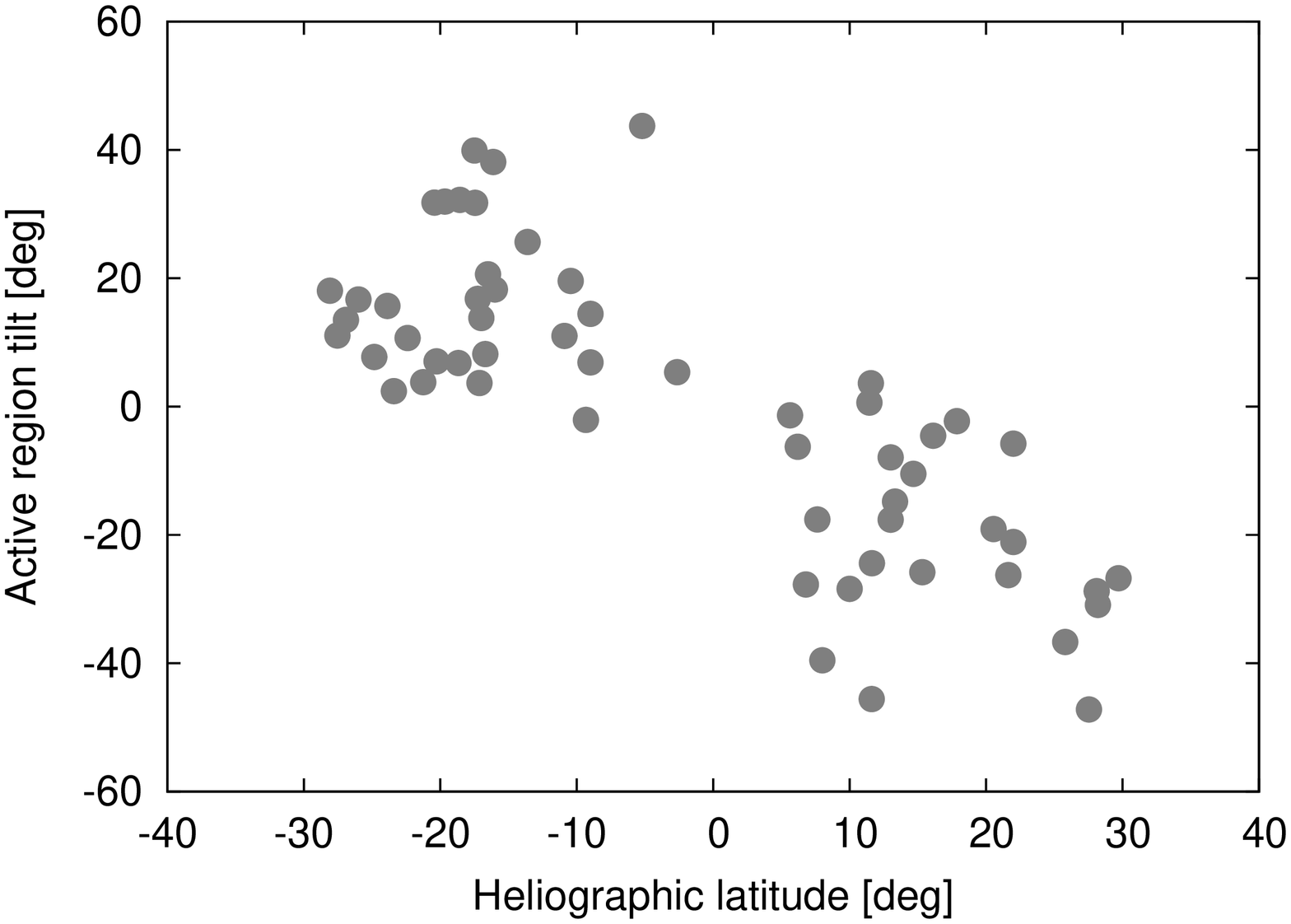}}}
\caption{The Joy's law in our sample of active regions. \emph{Left} -- calculated for the sample of active regions, which emerged in CMD less than 60 degrees. \emph{Right} -- the same for active regions, that emerged in CMD less than 60 degrees and spent at least 4 days within CMD less than 60 degrees, so due to the averaging in time the measured tilts are less noisy.}
\label{fig:joys}
\end{figure*}

\section{Data and Method}

In recent papers \citep[e.g.][]{2006AA...458..301S} we introduced a method to measure the large-scale dynamics in the solar photosphere. This method is based on supergranular-structures tracking in the full-disc, processed Dopplergrams, measured by MDI on-board the SOHO spacecraft. The application of this method allows one to compute the 24-hour averaged horizontal flow fields with resolution of 60\arcsec{} and noise level of 15~\mps. In the magnetised regions, this method makes it possible to measure the apparent motion of supergranular-scale magnetic features \citep{2009NewA...14..429S}.

The methodology consists of several steps. The main purpose of the pseudo-pipeline is to suppress disturbing effects, to remap the data onto a suitable coordinate system, to compute the horizontal vector-displacement field, and to convert the map of the displacements into the horizontal velocity vectors. The method and its subsequent validation are described in the great detail in \citet{2006AA...458..301S}. Here we provide a very brief overview.

The method processes 24-hour series of MDI full-disc Dopplergrams containing 1\,440 frames. The one-day series first undergoes noise substraction and the removal of other distorting effects (Carrington rotation profile, $p$-modes). Then, the processing of averaged frames consists of two main steps. The first step involves the calculation of mean zonal velocities using a very approximate \emph{local correlation tracking} \citep[LCT;][]{1986ApOpt..25..392N} algorithm. The obtained zonal velocities are fitted by a smooth function in a form of $\omega=c_0+c_1 \sin^2 b + c_2\sin^4 b$, and the series is tracked with the computed rotation profile. The tracking with the smooth velocity profile is done in order to minimise the average displacement of the supergranular structures caused by the solar differential rotation. Therefore the coefficients $c_0$, $c_1$, and $c_2$ vary from series to series. In the second step, the LCT algorithm, with an enhanced sensitivity, is applied to obtain a low-noise-level displacement map. Finally, the differential rotation (obtained in the first step) is added to the vector velocity field obtained in the second step. Both steps can be divided into several sub-steps, which do not differ much in application.

\begin{enumerate}
\item The data series containing 96 averaged frames is tracked using a selected rotation profile (Carrington rotation in the first step, differential rotation in the second step), and the frames are transformed into the Sanson-Flamsteed coordinate system to remove the geometrical distortion caused by the projection onto the disc. The Sanson-Flamsteed (also known as ``sinusoidal'') pseudo-cylindrical projection conserves the areas and is therefore suitable for the preparation
of the data used by LCT.

\item The tracked datacube undergoes $k$--$\omega$ filtering with the cut-off velocity of  1\,500~\mps{} to suppress the noise coming mostly from the evolutionary changes of the supergranules.

\item The LCT is applied: the lag between the correlated frames is 4~hours, the correlation window has a \emph{FWHM} of 60\arcsec, the measure of correlation is the sum of the absolute differences of the frames, and the nine-point method for calculation of the subpixel value of displacements is used. The calculated velocity field is averaged over the period of one day. The magnitudes of the vectors are corrected using formula 
\begin{equation}
\left( \begin{array}{c} v_{x,\rm corrected} \\ v_{y,\rm corrected} \end{array} \right) = 1.13 \left( \begin{array}{c} v_{x,\rm computed} \\ v_{y,\rm computed} \end{array} \right) - \left( \begin{array}{c} 15\ {\rm m\,s}^{-1}\\ 0 \end{array} \right)\ .
\end{equation}
\end{enumerate}
This formula results from the direct comparison between the synthetic data and the results of the displacement measuring procedure applied to these synthetic data \citep[see][]{2006AA...458..301S}.

The processed dataset contains 1004 full-disc flow maps in 502 days, when high-cadence MDI data were available and of a good quality. The velocity maps consist of many components on various spatial scales that cannot be reliably and unambiguously separated. On the largest scales, the differential rotation and the meridional circulation operate. We assume that these large-scale components vary slowly with the time, but their changes may influence the measurements of the proper motions of the active regions. Therefore, from each velocity map we subtracted the 13-day running average to remove the systematic changes in the flows on the largest scales. We assume that other components within our resolution of 60\arcsec{} are the components of our interest. The removal of the long-term average will also suppress any possible systematic errors. 

Based on the daily bulletins coming from the \emph{Space Environment Center of the National Oceanic and Atmospheric Administration (SEC NOAA)} we identified 564 labelled active regions in these maps. From which 522 were observed in more than one flow map. The theory predicts that the change in the dynamic regime should occur within a few (perhaps three) days after the magnetic regions has emerged in the photosphere. From the sample we therefore have to choose only those active regions that have emerged in the visible hemisphere within a reasonable distance from the central meridian, to avoid the influence of the edge effects. Only 194 active regions in the dataset were seen for the first time less than 60 degrees from the central meridian. This unfortunately means that the particular region could already be up to 1~day old at that time. Therefore, we included the measurements from the previous two days of the same region on the Sun into the analysis. From this set, we selected 72 bi-polar regions that survived at least 4 days within the central meridian distance (CMD) of less than 60 degrees. This drastic reduction of the sample was necessary in order to avoid the edge effects. Furthermore, to show that the disconnection-like behaviour is common among active regions showing significant changes in their dynamics over the lifespan, we chose another reduced sample of active regions which possessed a standard deviation of the rotation speed of more than 10~\mps{} over the lifespan. We have 18 such active regions in the sample. Note that the 10~\mps{} criterion is arbitrary.

The 24-hour averaged velocity maps and corresponding magnetograms typically contain more than one active region at once. Thus, the selection of individual active regions is needed. This selection is based on the masks applied to the full-disc displacement maps and magnetograms. The masks for each active region were obtained in a semi-automatic way. Firstly, from the daily reports from the \emph{SEC NOAA}, the location of the active region is determined and identified in MDI magnetograms. Then, for a wider area in the vicinity of the expected position of the active region, the structures of the magnetic field were roughly chosen by eye as the region-of-interest (closely corresponding with images from NASA's Solar Monitor, www.solarmonitor.org). The actual active region was identified in this region-of-interest by selecting pixels with a magnetic field strength above a threshold of 100~G. The 100-G region was then smoothed by 20\arcsec, which created the final mask. This mask was then applied to the set of displacement maps and magnetograms corresponding to days when the particular active region was recorded in the \emph{SEC NOAA} reports, and two days before.

For each active region in the sample of 72, we computed the properties suitable for a detailed investigation. To study the evolution of flows in the active regions and the properties describing these active regions, we calculated the large-scale flow field in the area covered by the magnetic field (with the threshold of 100 G), and the total unsigned flux. As additional parameters, we computed the mean magnetic field intensities in both the leading and trailing polarities, the coordinates of the gravity centre of both polarities, and the mean proper motions of both polarities in the frame co-rotating with the mean rotation rate of the active region. All these properties are sampled by 12 hours and cover the whole interval, when the particular active region was located within a CMD of 60 degrees. 

We searched for any significant changes in the dynamics, which could be caused by the disconnection of the magnetic structures from their deeper roots. The expected change should reveal itself as a sudden deceleration of the proper motion of the active region, because the radial gradient of the mean plasma rotation is negative in near-surface subphotospheric layers. All the active regions in the sample underwent this investigation in order to obtain a homogeneous set of active regions properties to examine.

\section{Results}

\subsection{Testing the sample}

The aim of this section is to show that the sample of active regions we have available displays the widely-accepted statistical properties of the magnetic field in the solar photosphere and is therefore suitable for more detailed analysis of the proper dynamics of active regions. We have selected two basic properties to test: the Joy's law and the polarity separation evolution.

As first noted by \citet{1919ApJ....49..153H}, the bi-polar sunspot groups are tilted with respect to the zonal direction. The latitudes of the leading polarities are on average lower than the latitudes of the trailing ones. The so-called \emph{Joy's law} shows that, statistically, the bi-polar regions are more tilted, if they are located at higher latitudes than the ones located at lower latitudes. The tilts are considered essential by kinematic dynamo models \citep[e.g.][]{1969ApJ...156....1L,1991ApJ...383..431W} and are often explained as the action of the Coriolis force on rising flux tube \citep[e.g.][]{1993AA...272..621D}. 

To test the Joy's law in our sample, we calculated the average magnetogram for each active region that emerged and remained in a CMD of less than 60 degrees. The tilt was measured from the centre-of-mass position of both polarities. The results are shown in Fig.~\ref{fig:joys}. We clearly see that the tilts are mostly positive on the southern hemisphere and negative on the northern hemisphere, i.e., the leading parts of the bi-polar active regions are closer to the equator than the following ones. Joy's law itself is not nicely recovered. However, as pointed out in \cite{2008ApJ...688L.115K}, Joy's law holds only statistically and we cannot expect much better results than those presented in Fig.~\ref{fig:joys}. Our results are very similar to those in \cite{2008ApJ...688L.115K}. Other effects, such as a dependence on the phase of the solar cycle as noted by \cite{2002ApJS..139..259U} may also take place. To separate such subtle effects the sample we have at our disposal is too sparse.

\begin{figure}[!]
\resizebox{0.49\textwidth}{!}{\rotatebox{270}{\includegraphics{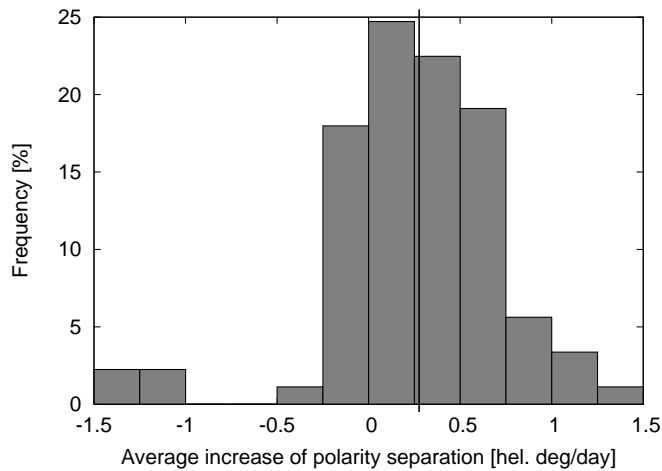}}}
\caption{The histogram of the polarity-separation speed for the sample of 194 active regions. It is evident that for most of the bi-polar active regions, the polarities are continuously separating. The characteristic (median) speed is overplotted by vertical line and it is roughly 0.3 heliographic degree per day.}
\label{fig:separations}
\end{figure}

Another property tested was the evolution of the separation distance between the leading and the following polarity in the spot group. Many works \citep[e.g][]{1990SoPh..126..285V,1994SoPh..153..449M,2007AA...472..277S} pointed out that during the evolution, the separation between the leading and the following polarities grows. The active region increases its size in the longitudinal direction. It is often interpreted as the result of the stretch caused by the differential rotation on the tilted region. The trailing polarity is at higher latitudes, therefore it senses a slower rotation than the leading part. We used our sample to measure this quantity. The separation of the leading polarity from the following one is defined as the distance between the gravity centres of both polarities in the Carrington coordinate system. We constructed the histogram of the separation speeds based on the sample of 194 active regions. The results are shown in Fig.~\ref{fig:separations}. We see that, on average, the separation of polarities in the active regions is increasing with time, with the median speed of 0.3 heliographic degree per day, which corresponds to $~$45~\mps{} on the solar equator.

\subsection{Acceleration or deceleration?}

As \citet{2004SoPh..221..225R} showed in a nice paper, based on the long-term observations of sunspots at Greenwich observatory covering more than a century, that rotation of sunspots usually slowed down during their evolution. In fact, this observation was initially shown by \citet{1962ZA.....55..110T}. These papers primarily deal with the deceleration of sunspots in term of linear relationship with time. However, there is no principal reason to assume that in a highly dynamic layer, such as is the convection zone, the deceleration should be monotonic. \citet{2003SoPh..214...65S}, using Kodaikanal data, found an opposite behaviour, i.e. that the rotation rates of spot groups increase with their age. They sorted the sunspot groups according to their life spans from one to eight days, and found that the rotation of the groups  on the first day of their life-cycle is the slowest. They then proceed to accelerate with age and reach the maximum velocity of rotation the day before dissolution.

\begin{figure}[b]
\resizebox{0.49\textwidth}{!}{\rotatebox{0}{\includegraphics{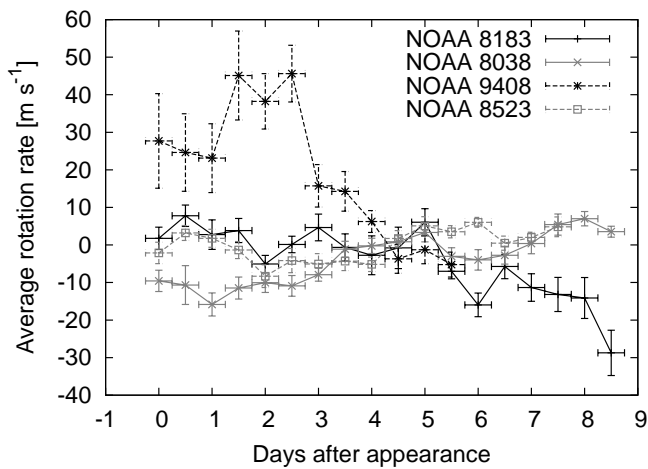}}}
\caption{Four different trends in the active region speed evolution, which are present in our data sample. The regions are continuously decelerating (NOAA~8183), or continuously accelerating (NOAA~8038). Less clear trends are also noticed, such as the early acceleration followed by a rapid deceleration (NOAA~9408) and very variable behaviour (NOAA~8523).}
\label{fig:evolutions}
\end{figure}

We studied the evolution of the proper velocities of the sunspots with respect to the averaged large-scale background, with a sampling of 12 hours. This sampling allowed for an improvement of the statistics compared to the simple average over the active region lifespan. Four different trends, which are present in the sample, are displayed in Fig.~\ref{fig:evolutions}. The error-bars represent the statistical errors showing the 1-$\sigma$ scatter of the velocities over the whole active region. 

\begin{figure*}
\centering
\resizebox{0.35\textwidth}{!}{\includegraphics{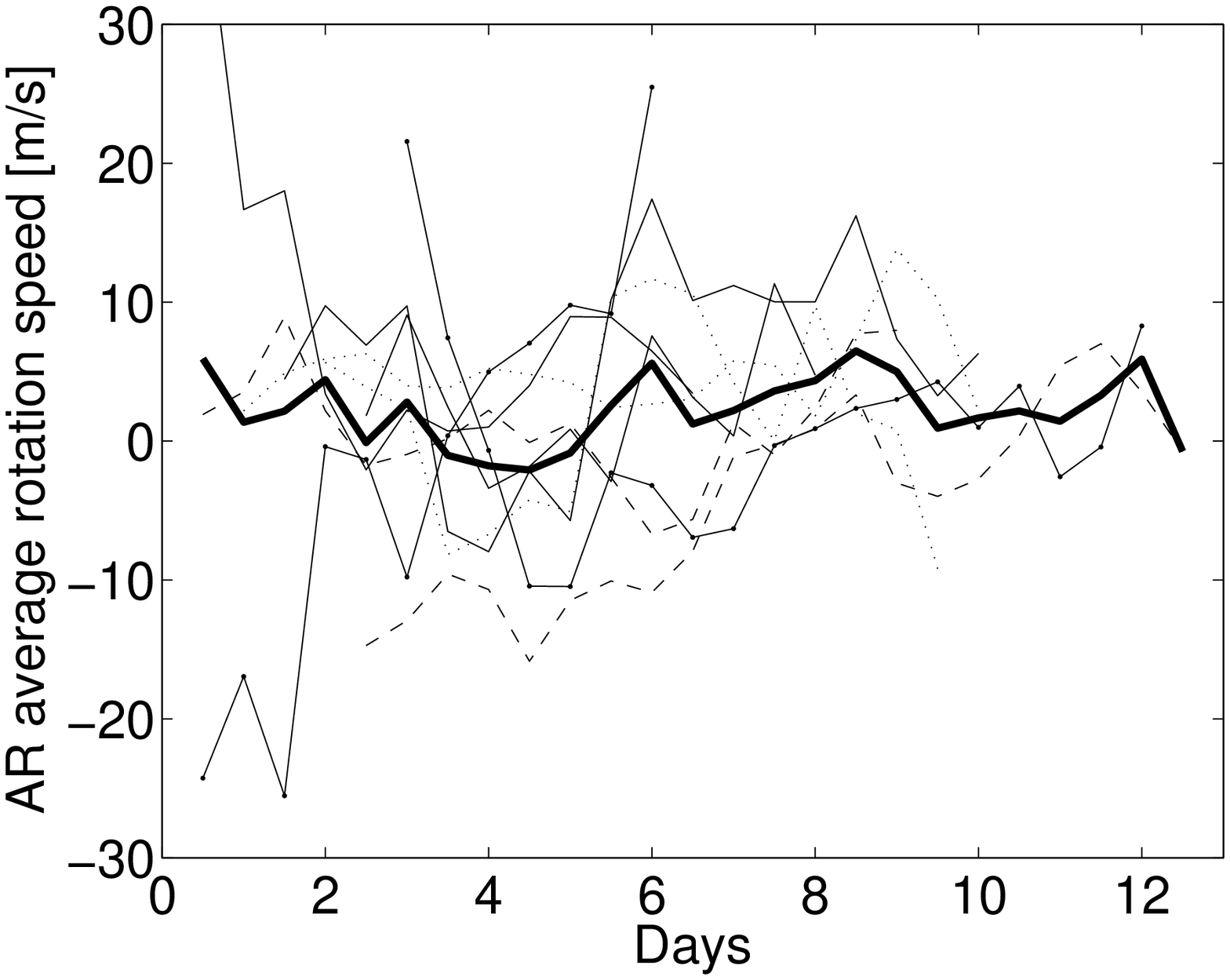}} \rule{7mm}{0pt}
\resizebox{0.35\textwidth}{!}{\includegraphics{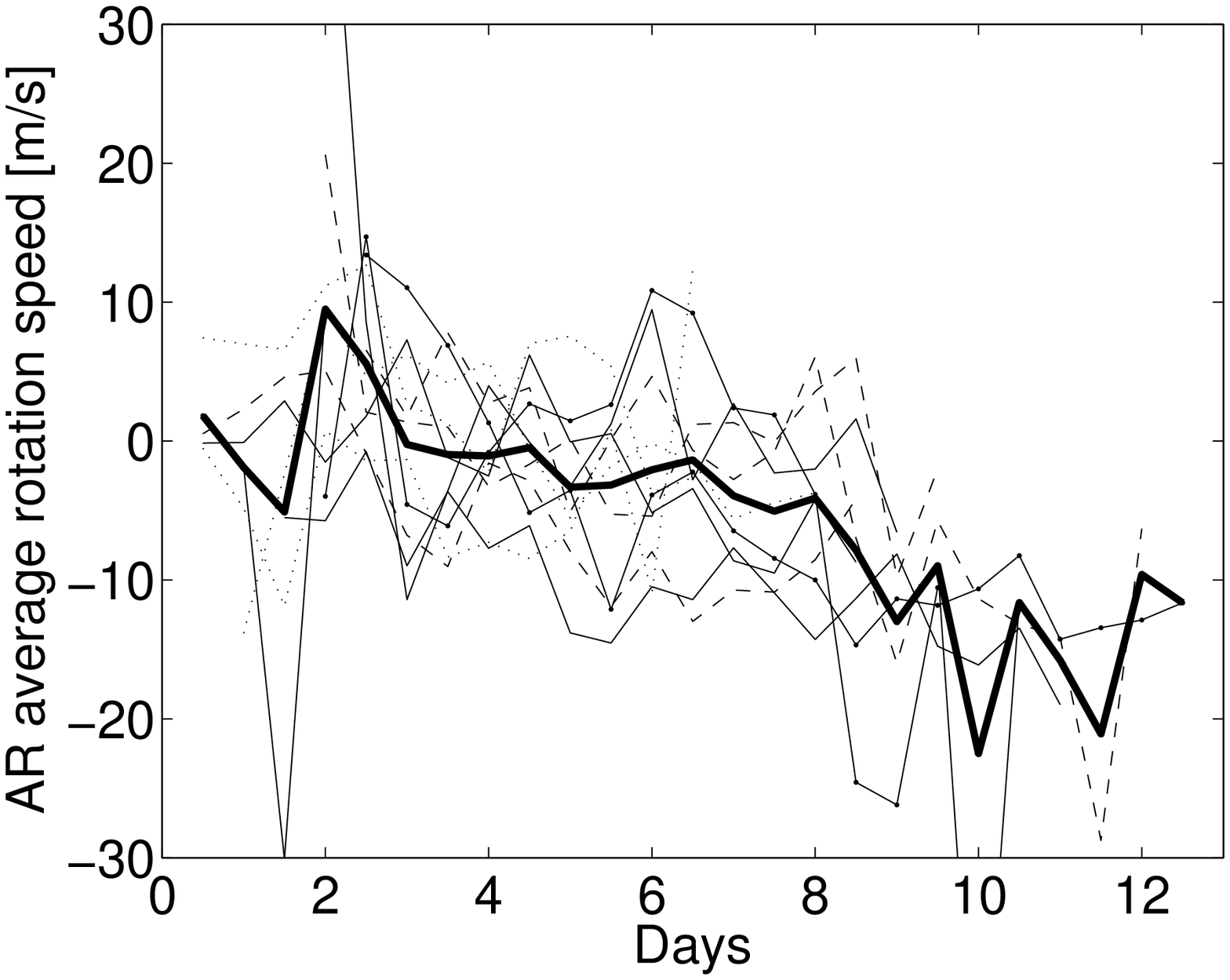}}\\
\resizebox{0.35\textwidth}{!}{\includegraphics{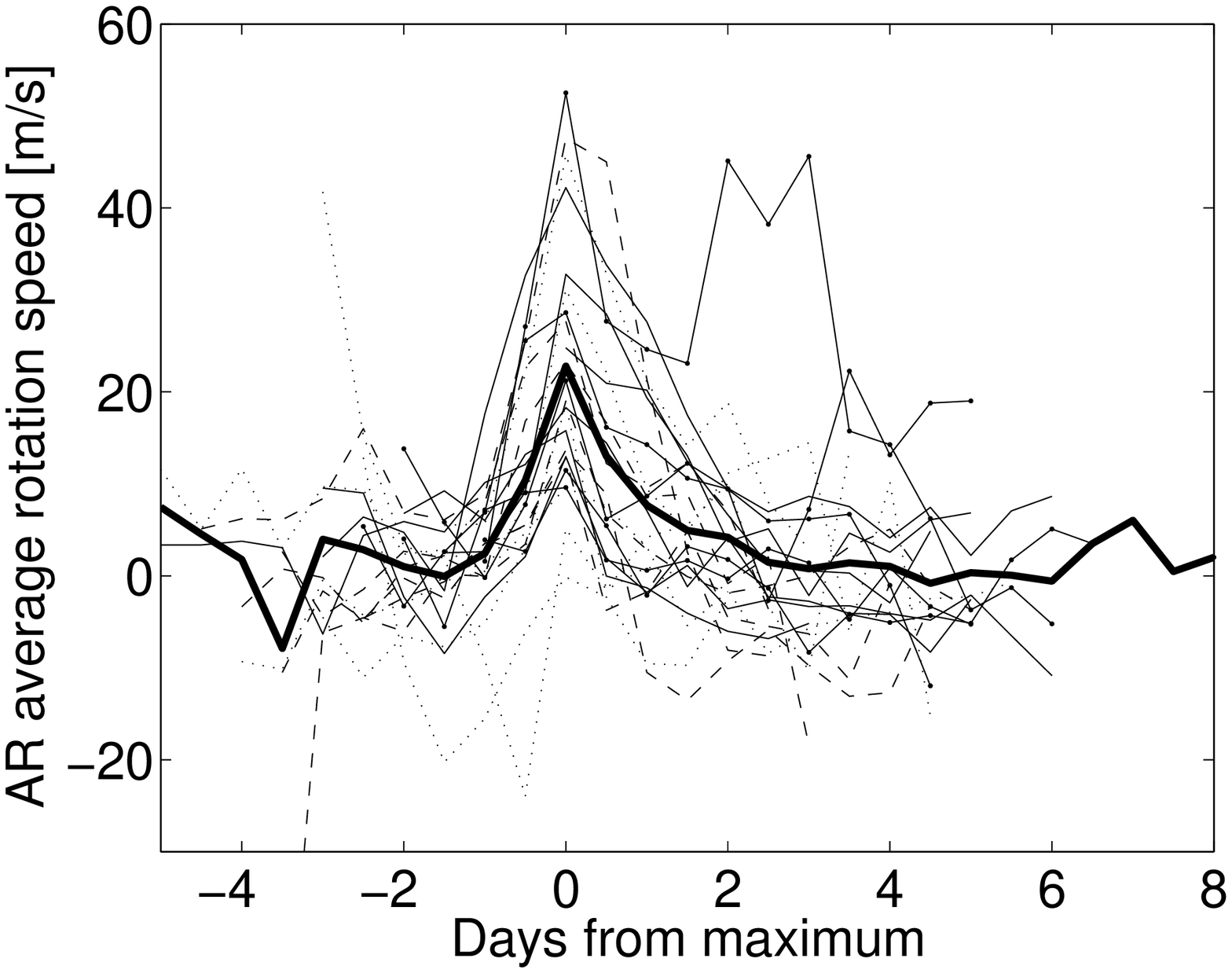}} \rule{7mm}{0pt}
\resizebox{0.35\textwidth}{!}{\includegraphics{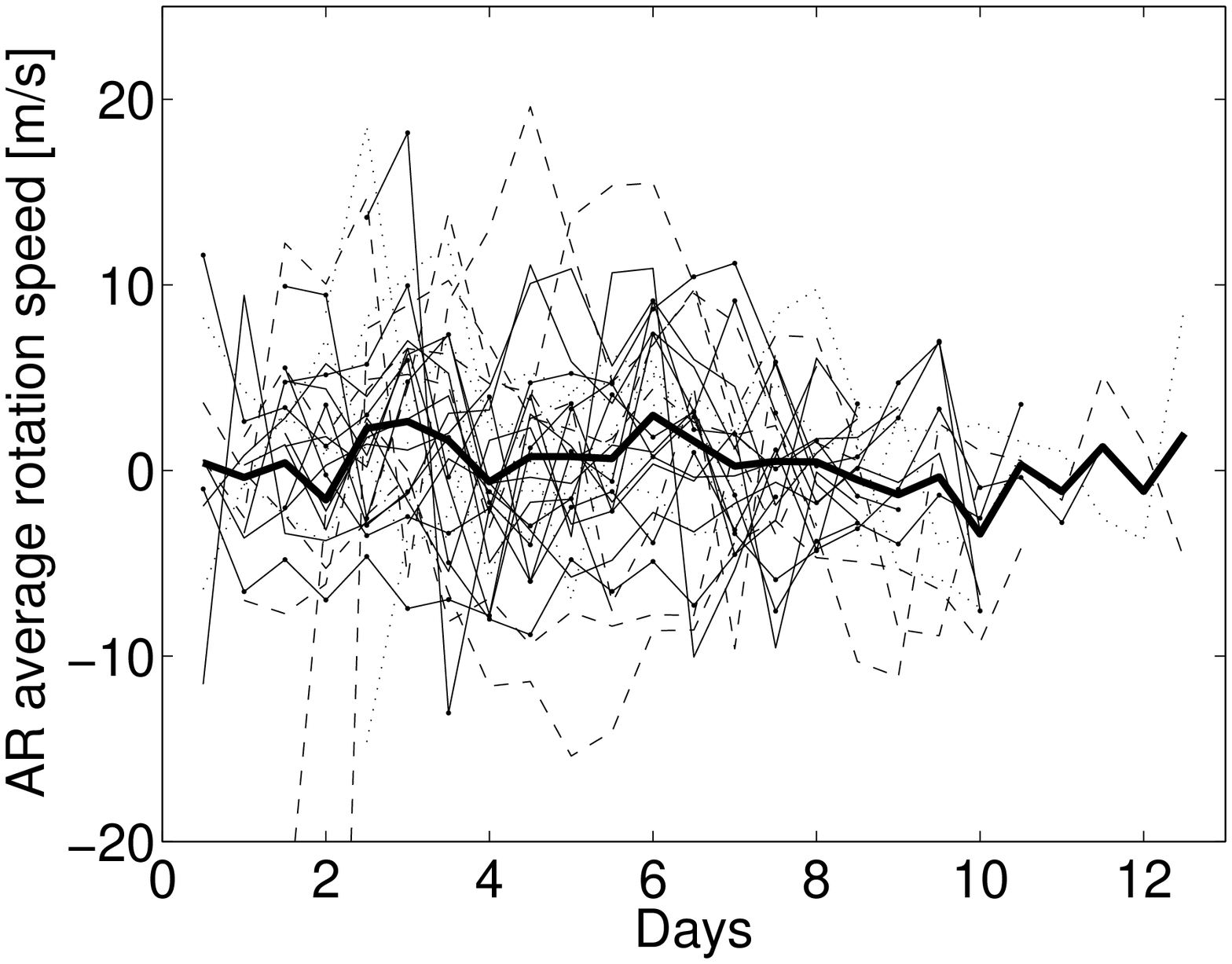}}
\caption{The division of the sample of the active regions into four dynamically different groups is based on the different shape of the trend of the active regions proper rotation speed with time. We divided 72 active regions in the group showing the continuous acceleration (upper left), continuous deceleration (upper right), acceleration followed by a sudden deceleration (bottom left), and the varying rotation speed (bottom right). With the thick line, the average trend among the displayed active regions is overplotted. Except for the third case, individual plots were not aligned. In the case with the possible dynamical disconnection imprint, we aligned the plots in time on the significant feature in the active regions lifetime, which is denoted by the one-to-two days lasting peak.}
\label{fig:groupstogether}
\end{figure*}

We observe that for two of these trends, i.e., continuous acceleration and variable behaviour, the change in the rotation speed is below 20~\mps. On the contrary, the other two cases display more significant changes. Table~\ref{tab:summary} contains the detailed results for all active regions in the sample. The sample is divided into four types based on the dynamic behaviour: The active region is continuously accelerating (\emph{A}), decelerating (\emph{D}), shows the signs of the dynamic disconnection (\emph{DC}), or behaves erratically (\emph{var}).  

The tabulated results show that 33~\% of the active regions in our sample first accelerate and then suddenly decelerate. In 13~\% of cases the active regions depict a continuous acceleration, while in 15~\% of cases a continuous deceleration is observed. Finally, in the remaining 39~\% a variable behaviour is present when the proper velocity does not change significantly or systematically in the active region lifespan. The active regions accelerate mostly during the first few days of their lifespan. Table~\ref{tab:summary} also contains a number of other values describing the characteristic properties of the active regions. These include the measure of the average slope of the deceleration or acceleration in time for active regions falling in the corresponding category, the variance of the rotation speed over the life-cycle of the active regions showing variable behaviour in their rotation speed. The measurements obtained for the active regions showing first acceleration and then deceleration can also be found in the same table, and are discussed in the following sections.

Given the number of active regions in the sample, it is not feasible to demonstrate the exact trends in proper rotation evolution for each active region we studied. Instead, we associate these trends to the above-mentioned corresponding groups, without making explicitly clear which plot belongs to which active region. These plots can be seen in Fig.~\ref{fig:groupstogether} and they demonstrate the overall trend in each selected group. Except for the third group, no alignment of the individual curves were done. The third group however contains a significant feature in the trend, a large peak lasting one to two days. The curves were therefore aligned in time using this feature. The thick solid line indicates the averages of all the trends contained in the panels. Using this average trend we are able to clearly distinguish different regimes for active regions showing continuous deceleration and for those representing the disconnecting group. Although it is difficult to clearly see the difference between the group with the assigned continuous acceleration and the variable behaviour, in the former case one can see the overall increasing rotation speed with time when following individual trends.

\begin{figure}[!b]
\resizebox{0.49\textwidth}{!}{\rotatebox{0}{\includegraphics{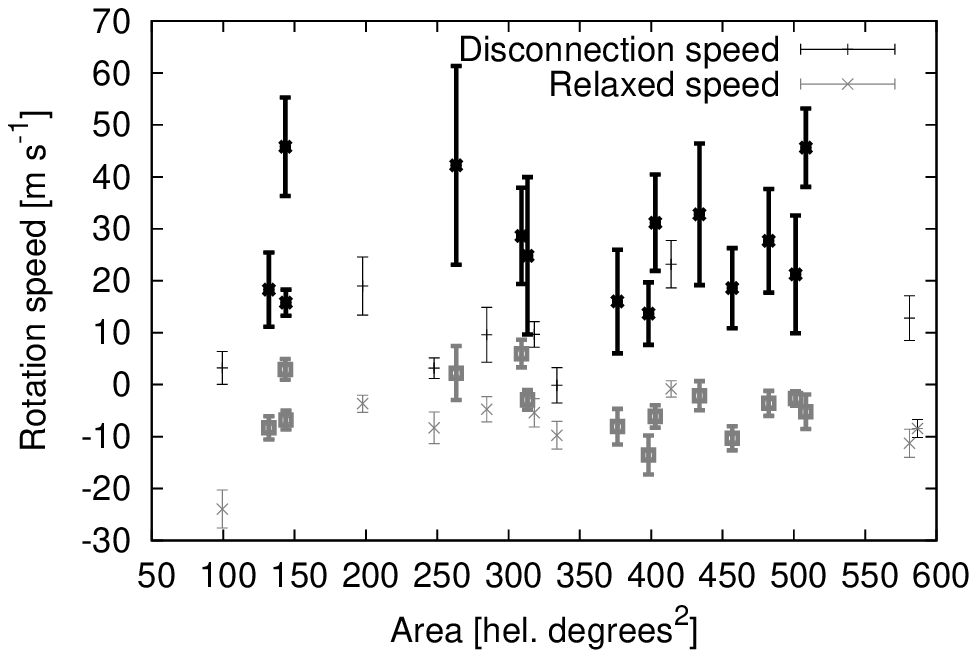}}}
\caption{The jump in the speed, which can be caused by the dynamic disconnection. For 23 active regions from the sample of 72 we see that the speed in the pre-disconnection phase is by some 50~\mps{} larger than in the post-disconnection phase. This value does not depend on the active region area. By the thick lines, active regions belonging to the subsample showing significant change in the speed during the lifespan is emphasised.}
\label{fig:max-min}
\end{figure}

\subsection{Possible evidence for disconnection}

From the sample, 18 active regions display a variance of speed evolution larger than 10~\mps, and from this 18, 78~\% are of the type which display an initial acceleration followed by a sudden deceleration. We assume that these active regions are representative of the sample which contain clear signatures of dynamic disconnection. 

\begin{figure*}[!t]
\centering
\resizebox{0.99\textwidth}{!}{\rotatebox{0}{\includegraphics{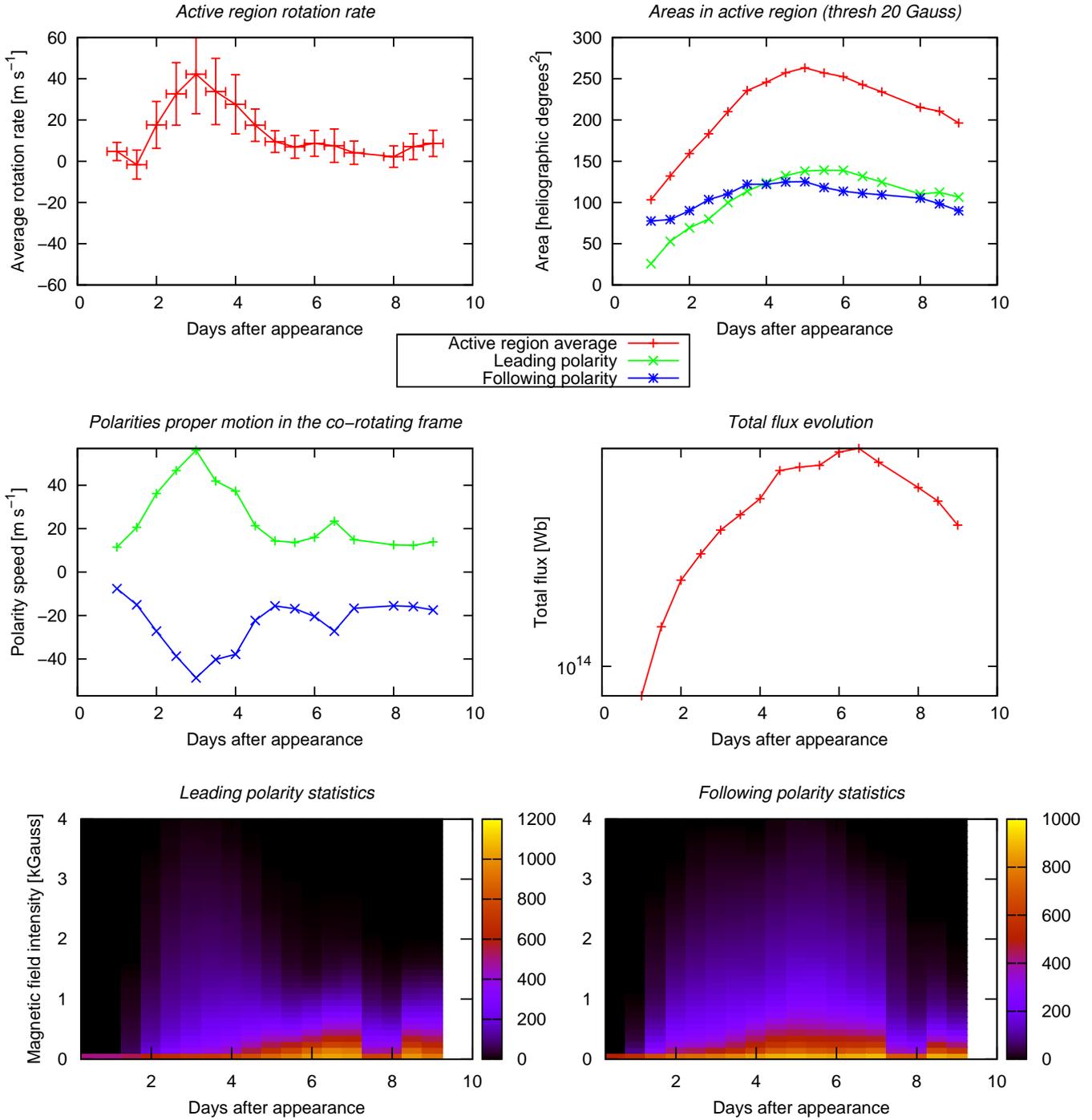}}}
\caption{The example of the evolution of flows, active region area, the total flux, and the histogram of the magnetic field intensities. We see the overall acceleration in first two days (in this phase, the polarities separate rapidly), and sudden deceleration after this instant. The area of the active region continues to grow for another two days, the total flux in the active region for three days (however, usually the maximum in the area corresponds with the maximum in the total flux). In the bottom row of figures, in each column the histogram of the magnetic field strength in the region-of-interest in the particular part is displayed in by colours. The evolution of the dynamics seems very symmetrical, which is not true for the histograms of the magnetic field strength. The actual evolution of the flows in this particular active region is displayed in Fig.~\ref{fig:example-flow-evolution}. }
\label{fig:example}
\end{figure*}

\begin{figure*}[!t]
\centering
\resizebox{0.65\textwidth}{!}{\rotatebox{0}{\includegraphics{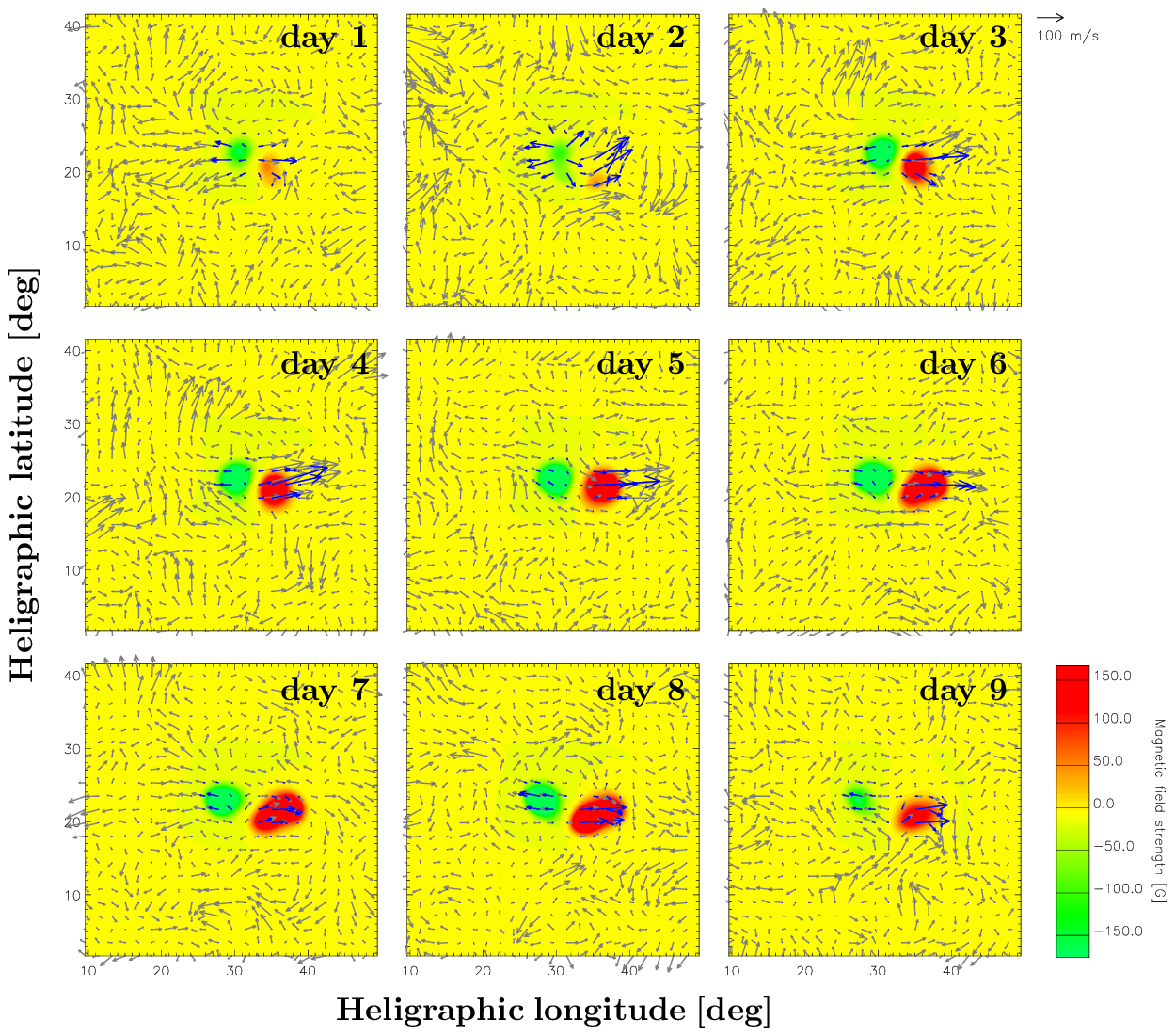}}}
\caption{The mosaic of the evolution of the flow field in the active region NOAA~8524. The background colours represent the large-scale magnetic field in the region, the grey arrows the the flows in the area. The differently coloured arrows in the magnetized areas represent the velocities that were used for the studies of the dynamics of this particular active regions.}
\label{fig:example-flow-evolution}
\end{figure*}

We assume that disconnection is exhibited as a sudden decrease in the proper speed, shortly before the maximum in area is reached. We further assume that the disconnection takes place at the instant of sudden decrease in speed. The upper part of the former flux-tube continues to rise and feeds more magnetic flux into the photosphere. Therefore the area of the magnetic field in the photosphere keeps growing until the disconnected part is fully emerged, after which the flux starts to disperse and the active region area decreases. In Fig.~\ref{fig:max-min} we display the speed at the assumed disconnection point (the \emph{disconnection speed}, $v_{\rm dis}$) and the speed in the following minimum (the \emph{relaxed speed}, $v_{\rm relax}$), these numbers are also summarised in Table~\ref{tab:summary}. The thick lines emphasise the set of 14 active regions which display a significant change in proper speed during their evolution, and which also exhibit the signs of disconnection. 

The typical drop of the speed during disconnection is $\sim$50~\mps. This value corresponds well with the bi-modal distribution of the equatorial rotation rates in the presence of sunspots (with velocities clustered around 1850~\mps{} and 1910~\mps), as found by \citet{2008AA...477..285S}. A quick check using the sunspot drawing archives showed that the ``fast'' group contained mostly the young and growing regions (which corresponds to the pre-disconnection phase in the current study), while the second ``scattered'' group contained mostly old and dispersing regions. In the current study this corresponds to the post-disconnection phase. 

Although our analysis is based on the mean rotation rate considered for the whole active region, the signatures of the disconnection should be detectable in both polarities. However, it is difficult to measure this phenomenon, because the motions of polarities in the co-rotating frame are an order of magnitude faster. The dynamics in the co-rotating frame (as displayed e.g. in Fig.~\ref{fig:example}) is usually very symmetrical in both polarities. Taking a careful look at this issue we found that in some 70~\% of active regions in the sample showing clear signatures of the disconnection, the disconnection features can be detected in both polarities, while in 30~\% it is clear only in one polarity. For a few cases, the time when disconnection occurs in opposite polarities differs by approximately 1~day. This issue is probably related to the configuration and asymmetry of the magnetic field in the particular active region. The disconnection is proposed for flux-tubes that form sunspots or pores, this does not apply to the magnetic field forming the plage. Unfortunately, as our method allows the measurement of large-scale features, we cannot address this issue properly, and it can possibly affect our results.

\subsection{Disconnection depth}

We observe a systematic shift $t_{\rm lag}$ between the time of the dynamic regime change and the time of the maximum in the area (see e.g. Fig.~\ref{fig:example}). We interpret this lag as the time which the flux tube forming the magnetic island rises from its parental magnetic structure to the surface after the disconnection. After the disconnection takes place, no more magnetic field is fed into the magnetic island and after its entire emergence the photospheric magnetic field starts to diminish. In all but two cases this lag is positive, therefore fulfilling the above mentioned assumption (see Table~\ref{tab:summary}).

\begin{figure}[!b]
\resizebox{0.49\textwidth}{!}{\includegraphics{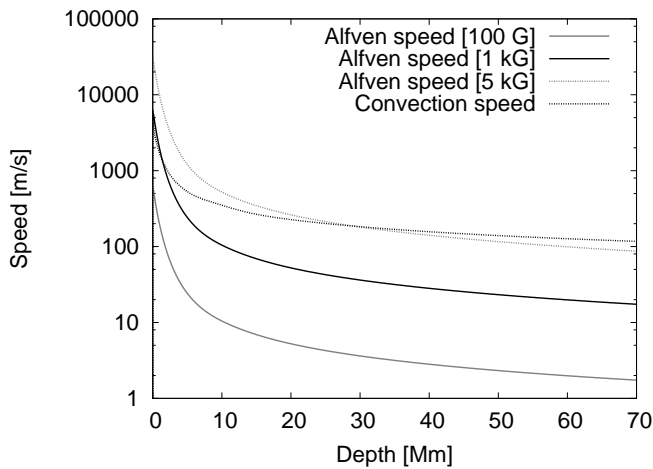}}
\caption{The trends in various speeds in the upper part of the solar convection zone. The Alfven speed and the convection speed are used to estimate the limits in depths, where the disconnection takes place. The computation of the speed trends is based on model S.}
\label{fig:speeds}
\end{figure}

The average lag is $1.6\pm 1.1$ days. The flux tube rises in the convection zone with a speed, $v_{\rm rise}$, which is unknown and must be determined for each case, e.g., by numerical simulations \citep[such as in e.g.][]{1993ApJ...405..390F}. We may roughly estimate this rising speed by various characteristic speeds in the convection zone, to put limits on the disconnection depth. Various speeds in the upper 70~Mm of the convection zone are displayed in Fig.~\ref{fig:speeds}.

The most reasonable estimate of the rising speed is the Alfven speed, $c_{\rm A}$, 
\begin{equation}
c_{\rm A}= \frac{B}{\sqrt{\mu \rho}}\ ,
\end{equation}
because the perturbations in the magnetic field usually propagate with this speed \citep[e.g.][]{1975ApJ...198..205P}. In the above equation $B$ is the intensity of the magnetic field, $\mu$ is the permeability, and $\rho$ is the plasma density. Based on the solar plasma parameters contained in the standard Model S \citep{1996Sci...272.1286C} atmosphere, we may estimate the rising time $t_{\rm rise}$ of the flux tube from the given depth in the convection zone $R_{\rm start}$ onto the surface: 
\begin{equation}
t_{\rm rise}= \int\limits_{R_{\rm start}}^{R_\odot} \frac{{\rm d}r}{c_{\rm A}(r)}=\int\limits_{R_{\rm start}}^{R_\odot} \frac{\sqrt{\mu \rho(r)}}{B}{\,{\rm d}r}\ .
\label{eq:depth}
\end{equation}

Assuming a constant 1-kG flux-tube emerging with the Alfven speed, the theoretical $t_{\rm lag}$ is 180 days if the flux-tube emerges from the bottom of the convection zone, and 7 days when emerging from the subsurface shear at 0.95~$R_\odot$. Even the $10^5$-G field would rise from the base of the convection zone for approximately one month. Although these numbers represent a very rough estimate, they show that a measured time-lag slightly more than 1~day corresponds to the disconnection location in the shallow subsurface layers. 

\begin{figure}
\resizebox{0.49\textwidth}{!}{\rotatebox{0}{\includegraphics{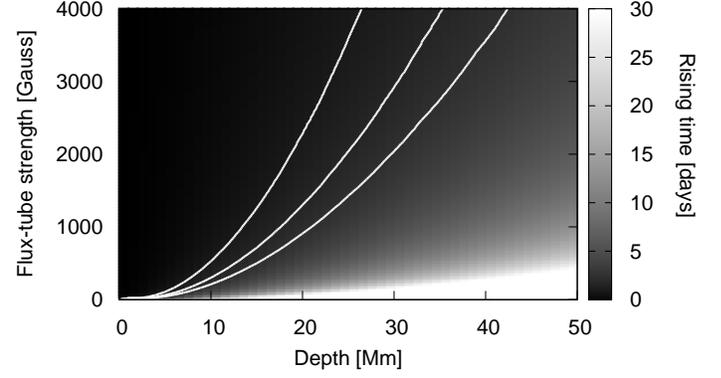}}}
\caption{The theoretical time-shift between the disconnection time and the time when the maximum in area is reached, based on the assumption that flux-tube rises with the Alfven speed as the function of depth and flux-tube strength. The measured time-shift including the error is displayed in contours. This very rough estimate shows that the disconnection cannot occur deeper than some 40~Mm for 4-kG flux-tube to reproduce the observed behaviour. The Alfven speed computation is based on Model S of \citet{1996Sci...272.1286C}. The real rising speed of an expanding flux-tube should not differ from the Alfven speed in the order of magnitude.}
\label{fig:lag_theory}
\end{figure}

In Fig.~\ref{fig:lag_theory} we show the theoretical rising time depending on the initial depth and the strength of the flux tube that is considered constant during the rising process. The measured lag is displayed in contours. We have to consider our theoretical results as an estimate, because the calculation does not include the flux-tube expansion, which naturally influences the rising speed. The precise modelling and the reproduction of the situation would be very difficult, because the initial properties of the flux tube deep in the convection zone are unknown. We believe that our estimate shows the essence of the problem. Inverting (\ref{eq:depth}) with $t_{\rm rise} \sim t_{\rm lag}$ we can estimate the disconnection depth for the particular active region. 

The Alfven speed gives the smallest disconnection depth estimate. Buoyant ascent of the flux tube requires the super-equipartition field, e.g., $c_{\rm A} > v_{\rm c}$, where $v_{\rm c}$ is a convection speed. Using the convection speed as the rising speed after the assumed disconnection in our study therefore provides the largest disconnection depth estimate. It is clear that the presence of the strong magnetic field alters convection and therefore probably changes the convection speed. However, the main purpose of this calculation is to bound the disconnection depth, therefore the use of the model convection speed without the influence of the magnetic field is justified. Again, without proper modelling it is not possible to go beyond presented estimates.

The mixing-length \citep[e.g.][]{1989sun..book.....S} convection speed in the Sun can be computed from a known solar model (e.g., Model S in our case) using the following formula:

\begin{equation}
v_{\rm c}=\sqrt{-\frac{Gm(r)}{r^2}A^*(r)\,\alpha^2 H_p(r)^2}\ ,
\end{equation}
where $m(r)$ is the total mass in the sphere with radius $r$, $\alpha$ is the mixing-length parameter, $H_p$ is a pressure scale height, and $A^*$ is a convection parameter, 
\begin{equation}
A^*(r)=\frac{1}{\gamma(r)} \frac{{\rm d} \ln p(r)}{{\rm d} \ln r}-\frac{{\rm d} \ln \rho(r)}{{\rm d} \ln r}\ ,
\end{equation}
where $\gamma$, $p$, and $\rho$ represent the adiabatic exponent, pressure, and density. The rise of the flux-tube with a convection speed implies that the strength of the field is the equipartition one. 

In Table~\ref{tab:summary} we provide the range of the disconnection depths for each active region studied. We observed that the disconnection occurs within a few tens of Mm in depth. The maximum intensity of the magnetic field used for the calculation of the Alfven speed is determined from the MDI magnetograms. We need to keep in mind that the actual strength of the original flux tube (also the rising time) may be different from the estimated one.

\section{Conclusions}

We conclude that the observed dynamical behaviour of a significant fraction of active regions in the homogeneous sample in the past solar cycle is compatible with a dynamic disconnection of the bi-polar active regions from their parental magnetic roots. In our sample, 33~\% of active regions display the signs of the disconnection. In the reduced sample with significant changes in the dynamics, 78~\% show the desired behaviour. The disconnecting active regions display a lag between the disconnection time and the instant when the maximum in area is established. If we assume that this lag is caused by the continuous rise of the disconnected magnetic field up to the photosphere, we may estimate the depth where disconnection takes place. We estimate that disconnection takes place within few tens of Mm below the solar surface, not in disagreement with 1-D simulations conducted by \citet{2005AA...441..337S}. 

The numerical simulations done by \citet{2005AA...441..337S} predict that the disconnection should happen for the significant fraction of bi-polar active regions, if not for all of them. However, our results show the clear attributes of the disconnection in only one third of the studied cases. We speculate that in other cases, disconnection does not occur during the observed span (e.g. before emergence into the photosphere in the case of continuously decelerating active regions) or that the influence of the proper motions of the magnetic features by the turbulent background is too high. Therefore we cannot reliably separate the attributes of the disconnection from other motions. 

The mechanism of the dynamic disconnection is proposed for strong flux-tubes that form sunspots or pores. With our investigation method we unfortunately do not have sufficient resolution to distinguish between the motion of magnetic structures in, e.g. sunspots, from the ones in the plage region, which probably embody different dynamic properties. 

Our present study does not provide a definitive answer on the question of whether the origin of magnetic activity is in the deep or shallow dynamo action. Both approaches require some form of disconnection of the surface magnetic field to allow the dispersed flux to migrate towards the solar poles. The deep dynamo based model assumed by \cite{2005AA...441..337S} provides predictions for particular active regions dynamics, which we have tested. Although the results of this test favour the model assumed, they are not robust enough to rule out other interpretations. The phenomenon of dynamic disconnection requires further investigation using high resolution methods and by local helioseismology, which proves its power in investigating deep structure of sunspots \citep[e.g.][]{2009ApJ...690L..72M}. Clear, deep magnetic field signatures would rule out a group of models assuming a shallow dynamo and provide other tests for rising flux-tubes models.

\begin{acknowledgements}
The authors were supported by the Grant Agency of Academy of Sciences of the Czech Republic under grant IAA30030808. The Astronomical Institute of ASCR is working on the Research project AV0Z10030501 (Academy of Sciences of CR), the Astronomical Institute of Charles University on the Research program MSM0021620860 (Ministry of Education of CR). SOHO is a project of international cooperation between ESA and NASA. The authors thank the referee Roger K. Ulrich, whose comments and suggestions significantly improved the quality of the study and its presentation, and also Hamed Moradi for correcting the English.
\end{acknowledgements}


\longtab{1}{
\begin{longtable}{l|cccc|cccccc}
\caption{\label{tab:summary} The studied properties of 72 active regions in our sample. Active regions of type \emph{DC} show signatures of dynamic disconnection.}\\

\hline\hline
 \bfseries NOAA & \bfseries From & \bfseries To & \bfseries Type & \bfseries Max. intensity & $\mathbf v_{\rm dis}$ & $\mathbf v_{\rm relax}$  &  \bfseries Slope & \bfseries Variance & \bfseries $t_{\rm lag}$ & \bfseries Disc. depth\\
 & &  & & Gauss & m\,s$^{-1}$ & m\,s$^{-1}$  &  m\,s$^{-1}$\,day$^{-1}$ & m\,s$^{-1}$ & days & Mm\\
\hline
\endfirsthead
\caption{continued.}\\

\hline\hline
 \bfseries NOAA & \bfseries From & \bfseries To & \bfseries Type & \bfseries Max. intensity & $\mathbf v_{\rm dis}$ & $\mathbf v_{\rm relax}$  &  \bfseries Slope & \bfseries Variance & \bfseries $t_{\rm lag}$ & \bfseries Disc. depth\\
 & &  & & Gauss & m\,s$^{-1}$ & m\,s$^{-1}$  &  m\,s$^{-1}$\,day$^{-1}$ & m\,s$^{-1}$ & days & Mm\\
\hline
\endhead


0651&2004/07/13&2004/07/22&var&$700$&  &  &  &$5.7$&  &  \\
0654&2004/07/24&2004/08/03&DC&$100$&$16(3)$&$-7(2)$&  &  &$-1.0$&
($5.0$--$26.7$)\\
0782&2005/06/27&2005/07/05&DC&$1300$&$14(6)$&$-14(4)$&  &  &$-1.5$&
($19.0$--$34.5$)\\
0783&2005/06/29&2005/07/09&DC&$1300$&$31(9)$&$-6(2)$&  &  &$-2.0$&
($21.9$--$41.5$)\\
0785&2005/06/28&2005/07/06&var&$800$&  &  &  &$6.5$&  &  \\
0791&2005/07/22&2005/07/31&A&$1400$&  &  &$0.7$&  &  &  \\
0793&2005/07/27&2005/08/04&DC&$700$&$29(9)$&$0(0)$&  &  &$-1.0$&
($11.7$--$26.7$)\\
7968&1996/06/01&1996/06/11&var&$200$&  &  &  &$3.7$&  &  \\
7978&1996/07/02&1996/07/12&var&$700$&  &  &  &$4.3$&  &  \\
8032&1997/04/14&1997/04/19&var&$500$&  &  &  &$4.7$&  &  \\
8038&1997/05/06&1997/05/16&A&$700$&  &  &$1.9$&  &  &  \\
8040&1997/05/15&1997/05/25&DC&$800$&$18(7)$&$-8(2)$&  &  &$-2.0$&
($17.2$--$41.5$)\\
8048&1997/05/29&1997/06/09&var&$1000$&  &  &  &$15.5$&  &  \\
8050&1997/06/08&1997/06/18&var&$300$&  &  &  &$4.3$&  &  \\
8052&1997/06/10&1997/06/19&A&$400$&  &  &$0.3$&  &  &  \\
8059&1997/07/02&1997/07/10&D&$200$&  &  &$-1.3$&  &  &  \\
8131&1998/01/10&1998/01/18&D&$1300$&  &  &$-1.7$&  &  &  \\
8141&1998/01/18&1998/01/26&var&$300$&  &  &  &$3.6$&  &  \\
8142&1998/01/20&1998/01/28&var&$800$&  &  &  &$4.2$&  &  \\
8145&1998/01/24&1998/02/02&A&$300$&  &  &$0.1$&  &  &  \\
8153&1998/02/06&1998/02/13&D&$400$&  &  &$-0.6$&  &  &  \\
8156&1998/02/11&1998/02/21&D&$1200$&  &  &$-2.2$&  &  &  \\
8160&1998/02/13&1998/02/20&var&$200$&  &  &  &$4.8$&  &  \\
8161&1998/02/16&1998/02/21&A&$300$&  &  &$0.7$&  &  &  \\
8164&1998/02/21&1998/03/01&A&$1000$&  &  &$-0.0$&  &  &  \\
8167&1998/02/22&1998/03/02&DC&$800$&$19(6)$&$-2(2)$&  &  &$-0.5$&
($9.0$--$17.4$)\\
8174&1998/03/05&1998/03/15&var&$1000$&  &  &  &$4.0$&  &  \\
8175&1998/03/07&1998/03/16&var&$200$&  &  &  &$6.7$&  &  \\
8180&1998/03/11&1998/03/21&D&$800$&  &  &$-1.1$&  &  &  \\
8183&1998/03/18&1998/03/28&D&$100$&  &  &$-2.4$&  &  &  \\
8184&1998/03/19&1998/03/26&var&$800$&  &  &  &$3.6$&  &  \\
8188&1998/03/25&1998/04/02&DC&$200$&$42(11)$&$-24(4)$&  &  &$-4.0$&
($12.4$--$64.9$)\\
8191&1998/03/30&1998/04/07&var&$500$&  &  &  &$6.6$&  &  \\
8193&1998/04/02&1998/04/09&var&$1800$&  &  &  &$4.0$&  &  \\
8507&1999/04/03&1999/04/12&var&$200$&  &  &  &$5.4$&  &  \\
8509&1999/04/05&1999/04/14&var&$500$&  &  &  &$8.0$&  &  \\
8511&1999/04/06&1999/04/16&var&$1300$&  &  &  &$4.2$&  &  \\
8512&1999/04/07&1999/04/17&var&$800$&  &  &  &$6.4$&  &  \\
8513&1999/04/08&1999/04/17&var&$800$&  &  &  &$4.1$&  &  \\
8515&1999/04/11&1999/04/19&D&$800$&  &  &$-1.2$&  &  &  \\
8523&1999/04/23&1999/05/02&DC&$300$&$3(2)$&$-8(3)$&  &  &$-3.0$&
($13.1$--$53.9$)\\
8524&1999/04/23&1999/05/02&DC&$1200$&$42(19)$&$2(5)$&  &  &$-2.0$&
($21.0$--$41.5$)\\
8526&1999/04/27&1999/05/07&var&$200$&  &  &  &$3.2$&  &  \\
8530&1999/05/01&1999/05/12&D&$500$&  &  &$-1.3$&  &  &  \\
8536&1999/05/05&1999/05/14&var&$1100$&  &  &  &$5.0$&  &  \\
9024&2000/05/28&2000/06/07&DC&$900$&$28(10)$&$-4(2)$&  &  &$0.0$&
($0.0$--$0.0$)\\
9028&2000/05/31&2000/06/08&var&$400$&  &  &  &$3.7$&  &  \\
9030&2000/06/03&2000/06/13&A&$500$&  &  &$-0.3$&  &  &  \\
9032&2000/06/03&2000/06/13&DC&$600$&$12(2)$&$-5(3)$&  &  &$-4.5$&
($22.3$--$70.0$)\\
9035&2000/06/07&2000/06/15&DC&$500$&$10(5)$&$-5(3)$&  &  &$-0.5$&
($7.4$--$17.4$)\\
9037&2000/06/09&2000/06/19&DC&$600$&$10(3)$&$-8(2)$&  &  &$-1.5$&
($13.1$,$34.5$)\\
9050&2000/06/15&2000/06/23&DC&$500$&$16(10)$&$-9(3)$&  &  &$-1.0$&
($10.0$--$26.7$)\\
9059&2000/06/21&2000/06/29&var&$500$&  &  &  &$5.9$&  &  \\
9063&2000/06/25&2000/07/05&var&$600$&  &  &  &$3.0$&  &  \\
9365&2001/03/01&2001/03/06&DC&$500$&$46(10)$&$0(0)$&  &  &$0.0$&
($0.0$--$0.0$)\\
9400&2001/03/23&2001/03/30&A&$400$&  &  &$5.3$&  &  &  \\
9401&2001/03/25&2001/04/03&D&$1300$&  &  &$-8.1$&  &  &  \\
9408&2001/03/27&2001/04/03&DC&$1600$&$38(7)$&$-4(4)$&  &  &$-1.5$&
($21.0$--$34.5$)\\
9424&2001/04/08&2001/04/16&DC&$700$&$33(14)$&$-2(3)$&  &  &$-1.0$&
($11.7$--$26.7$)\\
9426&2001/04/10&2001/04/18&var&$700$&  &  &  &$7.1$&  &  \\
9431&2001/04/16&2001/04/26&DC&$700$&$23(5)$&$-18(5)$&  &  &$-1.0$&
($11.7$--$26.7$)\\
9432&2001/04/16&2001/04/24&DC&$900$&$19(8)$&$-10(2)$&  &  &$-1.5$&
($15.9$--$34.5$)\\
9441&2001/04/26&2001/05/05&DC&$1200$&$21(11)$&$-3(1)$&  &  &$-2.5$&
($23.5$--$47.9$)\\
9442&2001/04/26&2001/05/06&D&$400$&  &  &$4.7$&  &  &  \\
9910&2002/04/17&2002/04/24&DC&$1200$&$25(15)$&$-3(2)$&  &  &$-2.5$&
($23.5$--$47.9$)\\
9915&2002/04/22&2002/04/29&var&$1500$&  &  &  &$3.5$&  &  \\
9922&2002/04/22&2002/05/01&A&$200$&  &  &$-2.0$&  &  &  \\
9932&2002/05/01&2002/05/11&var&$500$&  &  &  &$4.0$&  &  \\
9935&2002/05/01&2002/05/10&D&$500$&  &  &$0.4$&  &  &  \\
9936&2002/05/01&2002/05/10&DC&$400$&$13(4)$&$-6(3)$&  &  &$-1.0$&
($9.0$--$26.7$)\\
9950&2002/05/10&2002/05/19&var&$700$&  &  &  &$5.8$&  &  \\
9970&2002/05/26&2002/06/02&DC&$1500$&$0(3)$&$-10(3)$&  &  &$-0.5$&
($12.0$--$17.4$)

\end{longtable}
} 

\end{document}